\documentclass{ws-ijmpd}

\begin{document}


%
\catchline{}{}{}{}{}

\def\be{\begin{equation}}
\def\ee{\end{equation}}
\def\bea{\begin{eqnarray}}
\def\eea{\end{eqnarray}}
\def\nn{\nonumber}
\def\ep{\epsilon}
\def\c{\cite}
\def\m{\mu}
\def\ga{\gamma}
\def\lan{\langle}
\def\ran{\rangle}
\def\Ga{\Gamma}
\def\thet{\theta}
\def\la{\lambda}
\def\Lam{\Lambda}
\def\ka{\chi}
\def\si{\sigma}
\def\al{\alpha}
\def\pa{\partial}
\def\de{\delta}
\def\De{\Delta}
\def\Dex{\Delta_{,x}}
\def\Dey{\Delta_{,y}}
\def\Dev{\Delta^{-1}}
\def\Ome{\Omega}
\def\Om2{\Omega^{2}}
\def\ov{\over}
\def\gmn{g_{\mu\nu}}
\def\gmnv{g^{\mu\nu}}

\def\rsr{{r_{s}\over r}}
\def\rrs{{r\over r_{s}}}
\def\rs2r{{r_{s}\over 2r}}
\def\l2r2{{l^{2}\over r^{2}}}
\def\rsa{{r_{s}\over a}}
\def\rsb{{r_{s}\over b}}
\def\rsro{{r_{s}\over r_{o}}}
\def\rss{r_{s}}
\def\a2{{l^{2}\over a^{2}}}
\def\b2{{l^{2}\over b^{2}}}
\def\op{\oplus}
\def\sn{\stackrel{\circ}{n}}
\def\c{\cite}


\title
{ General description of Dirac spin-rotation effect with
relativistic factor}
\author{C. M. Zhang  }
\address{National Astronomical Observatories,
Chinese Academy of Sciences, Beijing 100012, China,
zhangcm@bao.ac.cn}


\maketitle

\begin{abstract}{\bf
The Mashhoon rotation-spin coupling is studied by means of the
parallelism description of general relativity. The relativistic
rotational tetrad is exploited, which results in the Minkowski
metric, and the torsion axial-vector and Dirac spin coupling will
give the Mashhoon rotation-spin term. For the high speed rotating
cases,  the  tangent velocity constructed by the angular velocity
$\Ome$ multiplying the distance r may exceed over the speed of light
c, i.e., $\Ome r \ge c$, which will make the relativistic factor
$\ga$ infinity or imaginary. In order to avoid this ``meaningless"
difficulty occurred in $\ga$ factor, we choose  to make the rotation
nonuniform and position-dependent in a particular way, and then we
find that the new  rotation-spin coupling energy expression
 is consistent with the previous results in the low speed limit.
 }
\end{abstract}




\section{Introduction}

{\bf On the rotation-spin coupling},
the phenomenon of rotation-spin coupling illustrates the inertia
of intrinsic spin, and its existence  was first proposed by
Mashhoon \c{mas74,mas88,mas99,mas20,tiom96}.  An experiment to
test for the existence of this term  has   been studied by
Mashhoon {\it et al.} and many others
\c{masexp1,masexp2,silv,sted97,lamb02}.
 It then follows easily \c{ryder92} that the coupling of spin with
rotation holds also at the relativistic level. It should be noted,
however, that this only holds in $\it Minkowski\;
spacetime$\c{rydermash}.

The straightforward theoretical derivations of the inertial effect
of Dirac particle have been performed  by Hehl and his
collaborators \c{hn,hehlexp1}, and these treatments have been
extended in several directions by a number of investigators
\c{huang,zcm01}, however, where the relativistic factor has not
been considered.  Hehl and Ni \c{hn} have calculated spin coupling
Hamiltonian in an arbitrary noninertial frame,  one subject both
to rotation and to acceleration, and rotation-spin coupling is a
result of their special case of zero acceleration. While, with regard to
 the existed references, we stress that the metric tensor
 implied from the exploited tetrad
by many authors \c{rydermash,hn,huang,tiom62,chap,ni78} is not a
Minkowski one, or the spacetime curvature exists, which seems
to declaim that the rotation produces the spacetime curvature.
Moreover,  there is no role of relativistic factor in their
 tetrads,  and   they have to constrain their discussions
within the light cylinder ($R_{L}=c/\Ome$) \c{rydermash} because of the
velocity limit of speed of light.  In principle, the description
of rotation-spin might  be in the special relativistic level and
 application of this effect could be  extended
anywhere in the rotational frame, within or beyond the light cylinder.

{\bf On the torsion-spin coupling}, it is one of most important characteristics
for the torsion gravity to compare with the general relativity (GR). As
a special version of torsion gravity, the parallelism of general
relativity (PGR) has been pursued by a number of authors
~\cite{hay79,nh80,perbook,per041,per042,mal79},
 where the spacetime  is characterized by the
 torsion tensor and the vanishing  curvature, named
Weitzenb\"ock spacetime~\cite{hay79}, which can be reduced from
the Riemann-Cartan spacetime where the generalized metric-affine
theory of
 gravitation is constructed \c{h76,hehl95,hehl99,per032}.
 PGR  will be equivalent to GR with
 the convenient choice of the parameters of  the Lagrangian \c{hay79},
however based on the tetrad  PGR is a powerful
and a natural tool to describe the Dirac field \c{per045}.

In notation of the symbols, we will use the Greek alphabet ($\mu$, $\nu$,
$\rho$,~$\cdots=1,2,3,4$) to denote tensor indices, that is,
indices related to spacetime. The Latin alphabet ($a$, $b$,
$c$,~$\cdots=1,2,3,4$) will be used to denote local Lorentz (or
tangent space) indices. The  tensor
and local Lorentz indices can be changed into each other with the
use of the tetrad $e^{a} {}_{\mu}$, which satisfy \be
e^{a}{}_{\mu} \; e_{a}{}^{\nu} = \delta_{\mu}{}^{\nu} \quad ;
\quad e^{a}{}_{\mu} \; e_{b}{}^{\mu} = \delta^{a}{}_{b} \; .
\label{orto} \ee A nontrivial tetrad field can be used to define
the linear Cartan connection\c{hay79,perbook}

\be
\Gamma^{\sigma}{}_{\mu \nu} = e_a{}^\sigma \partial_\nu e^a{}_\mu \;,
\label{car}
\ee
with respect to which the Cartan connection is
~\c{hay79,perbook}
\be
T^\sigma{}_{\mu \nu} =
\Gamma^ {\sigma}{}_{\mu \nu} - \Gamma^{\sigma}{}_{\nu \mu} \;.
\label{tor}
\ee
The metric is
\be
g_{\mu \nu} = \eta_{a b} \; e^a{}_\mu \; e^b{}_\nu \; ,
\label{gmn}
\ee
 where $\eta^{ab}$ is the metric in flat space with the line element

\be
d\tau^{2} = g_{\mu \nu} dx^{\mu} dx^{\nu} \;,
\ee
The irreducible torsion axial-vector can then be
constructed as~\c{hay79,perbook,pvz}
\be
A_{\m} = {1\over 6}\ep_{\m\nu\rho\si}T^{\nu\rho\si}\,,
\ee
with $\ep_{\mu \nu \rho \sigma}$ being the completely antisymmetric
tensor normalized as $\ep_{0123}=\sqrt{-g}$
and $\ep^{0123}=\frac{1}{\sqrt{-g}} $, where g is the determinant
of metric.

The spacetime dynamic effects on the spin is  incorporated into Dirac
equation  through the ``spin connection''  appearing in the Dirac equation
in  gravitation \cite{hay79,pvz}.
 In Weitzenb\"ock spacetime, as well as the general version of torsion
gravity, it has been shown by many
authors~\c{hay79,nh80,hehl71,tra72,rum79,yas80,aud81,ham94,ham95,ham02}
that the spin precession of a Dirac particle  is intimately
related to the torsion axial-vector,
 and it is interesting to note  that the
torsion  axial-vector represents the deviation of the axial symmetry
from the spherical symmetry \cite{nh80}.

\be
\frac{d{\bf S}}{dt} = - {3\ov2} \mbox{{\boldmath $A$}} \times {\bf S},
\label{precession1}
\ee
where ${\bf S}$ is the semiclassical spin vector of a Dirac particle,
and $\mbox{{\boldmath $A$}}$ is the spacelike part
of the torsion axial-vector.
Therefore, the corresponding extra Hamiltonian energy is of the form,

\be
\de H = - {3\ov2} \mbox{{\boldmath $A$}}\cdot \mbox{\boldmath$S$}\,.
\label{ham2}
\ee

Based on the torsion-spin coupling above, the Lense-Thirring
precession of Dirac spin has been obtained \c{pvz} by the standard
Kerr tetrad, and the rotation-spin precession can be achieved as
well \c{zcm01,zcm03}.

The purpose of the paper is constructing the connection between
the torsion-spin coupling and the rotation-spin coupling, and
especially we want to cope with the realistic physical conditions
 that  includes   the position outside the light
 cylinder ($R_{L}=c/\Ome$, see also \c{rydermash}) or
  the case of the tangent  velocity of the disk
exceeding over the speed of light c.
 For this end, the rotational tetrad is defined and the
resultant  metric is a Minkowski one, then the rotation-spin
 will be derived from the torsion-spin inferred from the
 systematical Dirac equation treatment. The motivation of the
 paper has two aspects, we firstly demonstrate that the rotation-spin
 can be described by the special torsion gravity, the parallelism
 description of general relativity, which possesses the deep
 theoretical foundations, and secondly the correct or effective
  description of
 rotation-spin will help us  recognize the new applications
 of torsion conception. On the latter, we may be confident of that
 we are extending the geometric-dynamics from the
   gravitation (Riemannian curved spacetime) to
 the rotation (Cartan torsional spacetime).

{\bf
 The organization of the paper is as follows: in section 2,
we discuss the rotation-spin coupling by introducing the
position-dependent angular velocity, and the new
 rotation-spin coupling formula is derived.
 The results and discussions  are presented
 in the last section. }


\section{The derivation of  rotation-spin effect}

Now we discuss the  Dirac equation in the
rotational coordinate system  with the polar  coordinates
$(t,r,\phi,z)$   with the rotating
angular velocity $\Ome$  set in z-direction.
 The tetrad can be expressed by the
dual basis of the differential one-form \c{hn,ho} through choosing
a coframe of the rotational coordinate  system \c{zcm03},

\bea \label{coframe0}
  d\vartheta ^{\hat{0}}& =& \,  \ga[cd\,t - (\Ome r/c) (rd\phi)]\;,
   \\
\quad d\vartheta ^{\hat{1}} &=&\, dr\;,  \\
 \quad d\vartheta ^{\hat{2}} &=&\,  \ga[ (rd\phi) -
(\Ome r/c)cdt]\;, \label{coframe}\\
 \quad d\vartheta ^{\hat{3}}&=& \, dz
\;, \eea where  the relativistic factor $\ga(r) = 1/\sqrt{1-(\Ome
r/c)^{2}}$.
 If $\Ome r $ is much less than the speed of light c,
 then we have  $\ga = 1 $  and the classical coframe expression is recovered,
which is same as those applied in the existed references
\c{hn,zcm01}. Therefore Eq.(\ref{coframe0}) and Eq.(\ref{coframe})
is a generalized relativistic coframe expression for any rotation
velocity, low or high energy cases.
    However, if $\Ome r/c > 1$, then the relativistic factor will
be meaningless, which will happen in the case of either $\Ome$ or r
being too large. {\bf Therefore, in order to overcome this
difficulty, we make choice of  a nonuniform and position-dependent
   angular velocity,   written as}
\be \Ome(r) = {\Ome \ov 1+(\Ome r/c)}\;. \label{omer} \ee Thus,  the
line velocity at radius r $v(r)=\Ome(r)r$ will not exceed over  the
speed of light. If $ r\rightarrow \infty$ or $ \Ome r\rightarrow
\infty$, we have $v(r) \rightarrow c$. $\Ome(r)<\Ome$ is implied
from Eq.(\ref{omer}) and  means that
 the  angular velocity  at the center of  rotating object is higher than that
 of far apart. While, Eq.(\ref{omer}) help us
 overcome the infinity difficulty in describing the line velocity of rotating system
 by considering the limited
 propagation  velocity in the non-rigid disk. The relativistic
factor is now modified as
 $\ga(r) = 1/\sqrt{1-(v(r)/c)^{2}}$.

The tetrad can be obtained with the subscript $\mu$ denoting the
column index (c.f. \c{hn,zcm01}) through replacing the original
constant $\Ome$ of $\Ome(r)$, with  the  conventional usage of the
unit of
 speed of light c=1,

\be e^{a}{}_{\mu} = \pmatrix{ \ga    & 0 &  -\ga \Ome(r) r^{2}& 0
\cr 0 & 1 & 0 & 0 \cr -\ga \Ome(r) r & 0 & \ga  r & 0 \cr 0      &
0 & 0 & 1} , \label{te1} \ee with its inverse

\be e_{a}{}^{\mu} = \pmatrix{ \ga   & 0  & \ga \Ome(r)  & 0 \cr 0
& 1 &0  &0\cr \ga \Ome(r) r &0 &\ga/r  &0 \cr 0 &0 &0  &1} .
\label{te2} \ee

We can inspect that the tetrad expressions Eqs.(\ref{te1}) and
(\ref{te2}) satisfy the orthogonal conditions in Eqs.(\ref{orto})
and (\ref{gmn}). From the tetrad  given above, we have the line
element and the implied metric, \bea d\tau^2 &=&
\eta_{ab}\,d\vartheta^{a} \otimes d\vartheta^{b}
= g_{\mu\nu}dx^{\mu}dx^{\nu} \nn \\
      &=&  dt^{2} - (dr^{2} + r^{2}d\phi^{2} + dz^{2}  )\;,
\label{dsr} \eea
and with the determinant of the metric
 \be
g=det|\gmn|= -r^{2}\;. \ee We find that the metric in
Eq.(\ref{dsr}) is a Minkowski one \c{mtw},  resulting in the null
Riemannian and Cartan curvatures. Although the curvature vanishes,
the torsion (field) may have  nonzero
 components because the torsion  is determined by the tetrad and  not by the metric.
 In other words, the basic element in the parallelism description of
 general relativity  is the
tetrad and the metric exploited is just a by-product
\c{hay79,perbook,per041,per042,h76,per032}. {}From Eqs.(\ref{te1})
and (\ref{te2}), we can now construct the Cartan connection, whose
components to contribute to the non-vanishing
 torsion axial-vectors are:
 \be
\Ga^{2}{}_{01} = - (\ga^{2}f) \Ome/r = - \Ome(r)/r,
\;\; \Ga^{0}{}_{21} = -(\ga^{2}f) \Ome r = - \Ome(r)r,
\ee
 with the factor definition \be f=1 - \Ome(r)r\; . \ee
 The corresponding non-vanishing torsion components
contributed to the   axial torsion-vectors  are:

\be T^{2}{}_{01} = -  \Ome(r)/r,   \;\; T^{0}{}_{21} =  -
 \Ome(r) r\;\;, \ee
and the  non-vanishing torsion axial-vectors  are consequently


\be A_{3} = {2\ov 3}\Ome(r) \; ,   \;\; A_{k} = 0,
k=0,1,2\;. \ee

As shown, $ A_{1}  = A_{2} = 0$ is  on account of the Z-axis
symmetry which results in the cancelling of the $r$ and $\phi$
components, and then generally we can write the torsion
axial-vector in the usual  vector expression $\mbox{{\boldmath
$A$}} = {2\ov3}{\bf \Ome}(r)$.
 From the spacetime geometry view,
the torsion axial-vector represents the deviation from the
spherical symmetry,  which will  disappear in
the spherical case (Schwarzschild spacetime for instance) and
occurs in the axisymmetry case (Kerr spacetime for instance) \c{nh80,pvz}.
 If the physical
measurement is performed in the rotating frame, the time dt is
taken as the proper time through setting the null space
difference.
Therefore the torsion axial-vector corresponds to an inertia field
with respect to Dirac spin, which is now clearly  expressed as a
precession equation by Eq.(\ref{precession1}),

\be \frac{d{\bf S}}{dt} = -  \mbox{{\boldmath $\Ome$}}(r)
\times {\bf S} \;. \label{precession2} \ee
The additive energy in the rotating frame is
\be \de H = -  \mbox{{\boldmath $\Ome$}}(r)\cdot
\mbox{\boldmath$S$}\;, \label{ham2} \ee
 which is a similar form  as that expected by Mashhoon (c.f. Ref.\c{mas88,mas20}) except
 the introduced varied angular velocity with the radius r
 for the sake of avoiding the infinite line velocity.

\section{Discussions and conclusions}

{\bf
 The obtained results of this paper are summarized
in the following:

  The Mashhoon rotation-spin coupling has been derived from the
 torsion-spin coupling by means of the parallelism of general relativity,
 where Dirac equation in the external field is treated by the
 introduced rotational tetrad. In the relativistic rotational tetrad,
 which has resulted in the flat metric, the relativistic factor
 $\ga$ is introduced, but it needs the tangent velocity of the
 observer to be less than the speed of light.
 In order to
 satisfy the need of the upper limit of velocity, we  choose
to make the rotation nonuniform  and position-dependent in a way as
described in Eq.(\ref{omer}).
 Then this choice makes the
  tangent velocity expression of rotating observer
  not exceed the speed of light at any
  positions, within or outside the light cylinder.

  On the description of the Mashhoon rotation-spin  coupling in
our parallelism version, the choice of   tetrad will produce a
Minkowski metric, a flat spacetime one, which arises  a new
conclusions on   the spacetime torsion and curvature. The rotational
field  can be an independent source for the production of torsion,
and even the gravitational field switches off. The existence of
torsion  in the flat spacetime, or the Minkowski spacetime,  can
help us   change the point of view that there is no torsion in the
flat spacetime \c{hay79,perbook,h76}.

  The extra additive energy of  the Mashhoon rotation-spin
coupling is expressed to be $\Ome(r)\hbar/2$, so in the low speed
limit of the present-day laboratory  $\Ome r \ll c$,  this coupling
energy is still $\Ome\hbar/2$ as obtained by Mashhoon before
\c{mas74,mas88,mas99}.  However, at the light cylinder of rotating
neutron star, $R_{L}=c/\Ome \sim 10^{7} (cm) (\Ome/500)^{-1}$, the
Mashhoon rotation-spin coupling will be half of that on the surface
of star, then we cannot sure  if  this effect can be detected in the
observations of the electromagnetic emissions in the rotating X-ray
neutron stars and black holes \c{klis04}. The quasi periodic
oscillations of X-ray flux have been discovered in many X-ray
neutron star and black hole systems since the launch of RXTE
satellite \c{klis04}, however these phenomena have not yet been
explained. Whether or not some
 oscillations of the X-ray energy  are involved in  the Mashhoon
 rotation-spin coupling still needs a thorough investigation.

}

\section*{Acknowledgments}
The author would like to thank  J. Pereira for discussions.
Thanks are due to   the critic comments from the anonymous referee
that greatly improve the  quality of the paper.


\begin{thebibliography}{10}


\bibitem{mas74} B. Mashhoon, {\it Nature}, {\bf 250}, 316 (1974);\\
B. Mashhoon, {\it Phys. Rev. D}, {\bf 10}, 1059 (1974);\\
B. Mashhoon, {\it Phys. Rev. D}, {\bf 11}, 2679 (1975).

\bibitem{mas88}
B. Mashhoon,  {\it Phys. Rev. Lett.} {\bf 61} 2639 (1988); {\bf
68}, 3812 (1992).
\bibitem{mas99}
B. Mashhoon, {\it Gen. Rel. Grav.}, {\bf 31},  681 (1999).

\bibitem{mas20}
B. Mashhoon, {\it Class. Quantum Grav.} {\bf 17}, 2399 (2000)

\bibitem{tiom96}
 I.D. Soares,  and J. Tiomno, {\it Phys. Rev. D} {\bf 54},
2808 (1996).

\bibitem{masexp1}
 B. Mashhoon,  {\it Phys. Lett. A}, {\bf 198}, 9 (1995).

\bibitem{masexp2}
B. Mashhoon, R. Neutze, M. Hannam, and G.E. Stedman, {\it Phys.
Lett. A} {\bf 249} 161 (1998).



\bibitem{silv}
 M. P. Silverman,   {\it Phys. Lett. A} {\bf 152}, 133 (1991);
 {\it Nuovo Cimento D}, {\bf 14}, 857 (1992).

\bibitem{sted97}
G.E. Stedman, {\it Rep. Prog. Phys.}, {\bf 60}, 615 (1997).



\bibitem{lamb02}
G. Papini, and G. Lambiase,  Phys. Lett. A294, 175 (2002)



\bibitem{ryder92}
L. Ryder, J.  Phys. A, {\bf 31}, 2465 (1992).

\bibitem{rydermash}
L. Ryder, and B. Mashhoon,  ``Spin and Rotation in General
Relativity",  gr-qc/0102101.





\bibitem{hn}
 F.W. Hehl, and W.T. Ni,
 {\it Phys. Rev. D} {\bf 42}, 2045 (1990)




\bibitem{hehlexp1} F.W. Hehl, J. Lemke, and  E.W. Mielke,  in
 {\it Geometry and Theoretical Physics}, edited by
 Debrus, J., and Hirshfeld, A. C. (Springer, Berlin), p. 56
 (1991);
 J. Audretsch,  F.W.  Hehl,  and C. L\"{a}mmerzahl, in
 {\it Relativistic Gravity Research}, edited by Ehlers,
 J., and Sch\"{a}fer, G. (Springer, Berlin), p. 368 (1992).


\bibitem{huang}
 J. Huang,  {\it Ann. Physik}, {\bf 3}, 53 (1994).



\bibitem{zcm01} C.M. Zhang,  and A.  Beesham {}
{\it Mod. Phys. Lett. } {\bf A16}, 2319 (2001).


\bibitem{tiom62}
C.G. de Oliveira, and J. Tiomno, {\it Nuovo Cimento} {\bf 24} 672
(1962.
\bibitem{chap}
T. C. Chapman,  and D.J. Leiter,
{\it Am. J. Phys.}, {\bf 44}, 858 (1976);\\
B. R. Iyer,  {\it  Phys. Rev. D} {\bf 26}, 1900 (1982);\\
E.  Schmutzer, and J. Pleba\'{n}ski,  {\it Fortschr. Phys.} {\bf
 25}, 37 (1977).


\bibitem{ni78}
W.T. Ni,  and M. Zimmermann, {\it  Phys. Rev. D} {\bf 17}, 1473
(1978).







\bibitem{zcm03}
C.M. Zhang,  {}  {\it Gen. Rel. Grav.} {\bf 35},  1465 (2003).




\bibitem{hay79}
K. Hayashi, and  T. Shirafuji, {\it Phys. Rev. D} {\bf 19}, 3524
(1979).

\bibitem{nh80}
J. Nitsch,  and  F.W. Hehl,  {\it  Phys. Lett. B}  {\bf 90},  98
(1980).








\bibitem{perbook}
R. Aldrovandi,  and J.G. Pereira,  {\it An Introduction to
Geometrical Physics},  (World Scientific, Singapore, 1995).




\bibitem{per041}
H.I. Arcos,  and J.G. Pereira,
 {\it Intl. J. of Mod. Phys. D}, D13, 2193, (2004),
 gr-qc/0501017.

\bibitem{per042}
H.I. Arcos,  V. C. de Andrade, and J.G. Pereira,
 {\it Intl. J. of Mod. Phys. D}, D13, 807, (2004),
  gr-qc/0403074.



\bibitem{mal79} J.W.  Maluf,
 {\it  Gen. Rel. Grav.} {\bf 31}, 173 (1979).











\bibitem{h76}
F.W. Hehl, P. von der Heyde, G. Kerlick  and J.M.  Nester, {\it
Rev. Mod. Phys.} {\bf 48},  393 (1976).



\bibitem{hehl95}
F.W. Hehl, J.D. McCrea, E.W. Mielke and Y. Ne'eman, Phys. Rep.
258, 1 (1995).

\bibitem{hehl99}
F.W. Hehl,  and A. Macias,   {\it Int. J. Mod. Phys. } {\bf D8},
399 (1999), gr-qc/9902076.

\bibitem{per032}
Yu. N. Obukhov,  and J. G. Pereira, Phys. Rev. D67, 044016 (2003),
gr-qc/0212080.


\bibitem{per045}
Yu. N. Obukhov, and J. G. Pereira,  Phys. Rev. D69, 128502 (2004).

\bibitem{pvz}
J.G. Pereira,  T. Vargas and C.M. Zhang, {\it
  Class. Quant. Grav.} {\bf 18},  833 (2001).





\bibitem{hehl71}
F.W. Hehl, Phys. Lett. A {\bf 36},  225 (1971).

\bibitem{tra72}
A. Trautman, Bull. Acad. Pol. Sci. Ser. Sci. Math. Astron. Phys.
{\bf 20}, 895 (1972).


\bibitem{rum79}
H. Rumpf, in {\it Cosmology and Gravitation}, eds. Bergmann, P. G.
and de Sabbata, V. (New York: Plenum), p63 (1980).

%



\bibitem{yas80}
P.B. Yasskin and R.W. Stoeger, {\it Phys. Rev.} {\bf D21}, 2081
(1980).

\bibitem{aud81}
J. Audretsch, {\it Phys. Rev.} {\bf D24}, 1470 (1981).



\bibitem{ham94}
R.T. Hammond, {\it Gen. Rel.  Grav.}  {\bf 26}, 247 (1994).

\bibitem{ham95}
R.T. Hammond, {\it  Cont. Phys.} {\bf 36}, 103 (1995).

\bibitem{ham02}
R. T. Hammond, Rep. Prog. Phys., 65, 599 (2002).


\bibitem{mtw}
Misner, C. W., Thorne, K. S. and Wheeler, J. A. 1973
                {\it Gravitation\/}
                (San Francisco: W. H. Freeman and Company);
                Weinberg S W 1972 {\em Gravitation and Cosmology}
                (New York: Wiley);
H.C. Ohanian and R. Ruffini,   {\it Gravitation and Spacetime \/}
(New York: Norton \& Company, 1994).



\bibitem{ho}
F.W. Hehl and Y.N. Obukhov, {\it Foundations of Classical
Electrodynamics, Charge, Flux, and Metric}, (Birkhauser, Boston,
2003).























































\bibitem{klis04}
M. van der Klis, Annual Rieview of Astron. \& Astrophys. {\bf  38},
717 (2000), (astro-ph/0001167); 
to appear in Compact stellar X-ray sources, eds.  W.H.G. Lewin \&
M. van der Klis, Cambridge University Press, (2004),
(astro-ph/0410551)


\bibitem{key-2}C. Chicone, B. Mashhoon and B. Punsly, Int. J. Mod. Phys. {\bf D13}, 945
(2004);\\
 C. Chicone and B. Mashhoon, Class. Quantum Grav. {\bf 19} 4231 (2002);\\
 C. Chicone and B. Mashhoon, Ann. Phys. (Leipzig) {\bf 14} 290 (2005).

\end{thebibliography}
\end{document}